\documentclass[pra,superscriptaddress,twocolumn,floatfix,showpacs,lengthcheck,nofootinbib]{revtex4-2}
\usepackage{graphicx}
\usepackage{dcolumn}
\usepackage{bm}
\usepackage{hyperref}

\usepackage[utf8x]{inputenc}
\usepackage[english]{babel}
\usepackage[T1]{fontenc}
\usepackage{bbold} 
\usepackage{amsthm}

\usepackage{soul} 

\usepackage[caption=false]{subfig}

\usepackage{braket}
\usepackage{cancel}
\usepackage{xcolor}
\usepackage{stackrel}
\usepackage{mathtools}

\newcommand{\Hil}{\mathcal{H}}  
\newcommand{\E}{\mathcal{E}}    
\newcommand{\A}{\bm{a}}   
\newcommand{\q}{\bm{q}}   
\newcommand{\p}{\bm{p}}   

\newcommand{\Tr}[1]{\mathrm{tr}\left[#1\right]}          
\newcommand{\partTr}[2]{\mathrm{tr}_{#1}\left[#2\right]}   

\newcommand{\sys}{\mathrm{S}}
\newcommand{\bath}{\mathrm{B}}
\newcommand{\inter}{\mathrm{I}}
\newcommand{\total}{\mathrm{Tot}}

\newcommand{\dis}{\mathrm{d}}  

\newcommand{\1}{\mathbb{1}}   
\newcommand{\ii}{\mathrm{i}}   
\newcommand{\e}{\mathrm{e}}  

\newcommand{\kbT}{k_\mathrm{B} T}  
\newcommand{\Trace}{\mathrm{tr}}   
\newcommand{\Kol}{\mathrm{kol}}    

\newcommand{\with}{\mathrm{with}}
\newcommand{\und}{\mathrm{and}}

\newcommand{\sigmabm}{\bm{\sigma}}  
\newcommand{\vecmu}{\vec{\mu}}      
\newcommand{\Omegabm}{\bm{\Omega}}  
\newcommand{\G}[2]{\bm{G}^{#1}_{#2}}           
\newcommand{\Gauss}[2]{\mathcal{G}^{#1}_{#2}}  
\newcommand{\x}{\vec{x}}  
\newcommand{\y}{\vec{y}}   
\newcommand{\X}{\vec{\bm{X}}}    

\newcommand{\br}[1]{\left(#1\right)}
\newcommand{\Br}[1]{\left[#1\right]}

\newcommand{\ketbra}[2]{|#1\rangle\langle#2|} 
\newcommand{\Rho}{\bm{\rho}}  
\newcommand{\F}{\mathcal{F}}  
\newcommand{\sopt}{\xi}    
\newcommand{\ubs}[2]{\stackrel[#1]{}{\underbrace{#2}}}  

\newtheorem*{definitions*}{Definitions}

\begin{document}

\title{Phase space measures of information flow in open systems: 
A quantum and classical perspective of non-Markovianity}

\author{Moritz F. Richter}

\affiliation{Institute of Physics, University of Freiburg, 
Hermann-Herder-Stra{\ss}e 3, D-79104 Freiburg, Germany}

\author{Heinz-Peter Breuer}

\affiliation{Institute of Physics, University of Freiburg, 
Hermann-Herder-Stra{\ss}e 3, D-79104 Freiburg, Germany}

\affiliation{EUCOR Centre for Quantum Science and Quantum Computing,
University of Freiburg, Hermann-Herder-Stra{\ss}e 3, D-79104 Freiburg, Germany}

\vspace{10pt}

\begin{abstract}
The exchange of information between an open quantum system and its environment, especially the backflow of information from the environment to the open system associated with quantum notions of non-Markovianity, is a widely discussed topic for years now. This information flow can be quantified by means of the trace distance of pairs of quantum states which provides a measure for the distinguishability of the states. The same idea can also be used to characterize the information flow in classical open systems through a suitable distance measure for their probability distributions on phase space. Here, we investigate the connection between the trace distance based quantum measure and the Kolmogorov distance for differently ordered quasi-probability distributions on phase space. In particular, we show that for any pair of quantum states one can find a unique quasi-probability distribution for which the Kolmogorov distance coincides with the trace distance. We further study the quantum-to-classical transition of the distance measures. Employing the Caldeira-Legget model of quantum Brownian motion as a prototypical example, numerical simulations indicate a particularly rapid convergence of the Kolmogorov distance of the Wigner functions to the trace distance in the classical uncertainty limit, which establishes the Wigner function distance as an optimal tool for measuring semi-classical information backflow and for quantifying non-Markovianity in open continuous variable quantum systems.
\end{abstract}

\maketitle

\section{Introduction}
\label{Introduction}
Although in many applications closed quantum systems completely shielded from their environment are desirable, it is necessary for any realistic theory to also include effects caused by interaction with said environment -- often assumed to be some kind of thermal bath. Yet such a theory of \textit{open quantum systems} \cite{Breuer-Petruccione_2007, Gardiner-Zoller_2010} does not only serve as necessary complication but also offers a wide range of interesting features giving additional insight into quantum mechanics such as the theory of decoherence \cite{Zurek2003, Schlosshauer2019} or quantum thermodynamics \cite{QuantumThermo2018,Landi2021}. Another aspect, the present letter will be concerned with, is the exchange of information between a system and its surrounding bath often associated with notions of quantum 
non-Markovianity \cite{Breuer2009,Laine2010,Laine2010b,Vacchini2011,Rivas2014,Breuer2016,Li2018}. Information flow is not only an interesting topic on its own but also of obvious importance in realistic applications of quantum computation and  quantum technologies in general \cite{Tancara2023,Bylicka2014}. Thus, to quantify the exchange of information reveals not only features of certain quantum processes at hand but can also be a tool of benchmarking said applications.

To quantify information flow one can track measures of distinguishability between two initial states through time (see Sec.~\ref{Measures of Information Flow} below). Such measures can be, for example, suitable distances like the trace distance \cite{Breuer2009,Laine2010} or the more general trace norm of the Helstrom matrix \cite{Chruscinski2011,Wissmann2015} as well as entropic quantities \cite{Settimo2022} of the density operators for which one accordingly needs to solve the time evolutions under the given dynamics of the system. For high dimensional systems such computations can be challenging. Additionally, if one prefers to describe states of the system by some other means than density operator one would also seek for measures of distinguishability working directly with the chosen representation of states instead of computing the density operator first. One such example are continuous variable quantum (CVQ) systems as used, for example, in quantum optics \cite{Scully-Zubairy_2006,Serafini_2017}, where one naturally describes states by quasi-probability distributions over phase space instead of density operators on Hilbert spaces. Moreover, information flow between system and environment is not at all a genuine quantum feature, but should be present also in classical systems as well.

In this letter we construct a suitable measure for the information flow in classical phase space models which is based on the Kolmogorov distance between quasi-probability distributions on phase space and study its relation to the quantifier based on the trace distance between quantum states. Furthermore, we show that in a specific semi-classical limit of mixtures of Gaussian states the measures of the information flow based on the classical Kolmogorov distance and the quantum trace distance converge to each other. These results are illustrated by means of the Caldeira-Legget model of quantum Brownian motion \cite{Caldeira1983}. For this standard model of open system dynamics we demonstrate that the Kolmogorov distance of the Wigner functions converges especially fast to the quantum trace distance, suggesting it as a kind of optimal quasi-probability distribution measuring the information flow.

The paper is organized as follows. In Sec.~\ref{Measures of Information Flow} we briefly recapitulate the mathematical description of open quantum systems, their dynamics and how to quantify the flow of information between the open system and its environment in terms of the distinguishability of pairs of quantum states. We also translate this concept to classical phase space models by constructing a suitable distinguishability measure using the Kolmogorov distance between phase space distributions. In Sec.~\ref{Continuous Variable Quantum Systems} we study continuous variable quantum systems and introduce the class of s-ordered quasi-probability distributions on phase space uniquely characterizing quantum states. In Sec.~\ref{Quantum-to-Classical Transition} we then explain the relation between the quantum and the classical measures of information flow, as well as the transition from one to the other for increasing uncertainty, i.e. width in phase space, of the states. An exemplary application and visualization of this transition is discussed in Sec.~\ref{Application to the Caldeira-Leggett model}, employing the Caldeira-Leggett (CL) model of quantum Brownian motion. Finally we summarize our findings and draw some conclusions in Sec.~\ref{Conclusion}.

\section{Measures of Information Flow}
\label{Measures of Information Flow}

\subsection{Open quantum systems}
\label{Open quantum systems}
We consider an open quantum system S with Hilbert space $\Hil_\sys$ coupled to a bath B with Hilbert space
$\Hil_\bath$. The Hilbert space of the total system is given by the tensor product 
$\Hil_\total \ = \ \Hil_\sys \otimes \Hil_\bath$, and the Hamiltonian of the total system can be divided into three parts as
	\begin{align}
	\bm{H}_\total \ = \ \bm{H}_\sys \otimes \1_\bath + \1_\sys \otimes \bm{H}_\bath + \bm{H}_\mathrm{int},
 	\end{align}
where the first two terms describe the Hamiltonian of the system S and of the bath B, respectively, while the third
term represents the system-bath interaction. 
The state of the total system is represented by a density matrix 
$\Rho_\total$ which evolves unitarily according to
	\begin{align}
	\Rho_\total(t) \ &= \ \e^{-\frac{\ii t}{\hbar} \bm{H}_\total} \ \Rho_\total \ \e^{+\frac{\ii t}{\hbar} \bm{H}_\total}.
	\end{align}
The open system's density matrix is then obtained by the partial trace
	\begin{align}
	\Rho \ := \Rho_\sys \ = \ \partTr{\bath}{\Rho_\total}.
	\end{align}
As suggested by this last equation, omitting the index by default we will refer to the open system.
Assuming a product state of system and bath at the initial time, $\Rho_\total (0) = \Rho (0) \otimes \Rho_\bath$, one can describe the evolution over some time interval $\mathbb{T}$ by means of
	\begin{align}
	\begin{split}
	\Rho(t) \ &= \ \Lambda(t) \Br{\Rho(0)}
	\\
	&= \ \partTr{\bath}{\e^{-\frac{\ii t}{\hbar} \bm{H}_\total}  \br{\Rho(0)\otimes\Rho_\bath} \e^{+\frac{\ii t}{\hbar} \bm{H}_\total} },
	\end{split}
	\label{eq:Stinespring}
	\end{align}
defining a family $\{\Lambda(t)\}_{t\in\mathbb{T}}$ of quantum channels or completely positive and trace preserving dynamical maps \cite{Nielsen-Chuang_2000}. 

The interaction between system and bath leads to a transfer of information from the system to the environment and back from the environment to the system \cite{Breuer2009}. This information is encoded in the system or environmental degrees of freedom, or in the correlations between system and environment \cite{Laine2010b,Breuer2016}. Relating the information encoded in the open system degrees of freedom by means of the distinguishabilty of quantum states, one can quantify the flow of  information between system and bath by tracking a suitable measure of distinguishability through time. Any decrease of the distinguishability then signals information flowing from the system to the bath which can be interpreted as a loss of information from the system and, hence, as Markovian dynamics. Vice versa, any increase of the distinguishabilty signifies information flowing from the bath back to the system implying a gain of information and non-Markovian dynamics (memory effects) \cite{Breuer2009,Laine2010}. 

A suitable measure for distinguishability is the trace distance defined by
\begin{align}
 \dis_\Trace\br{\Rho_1, \Rho_2} \  :=  \frac{1}{2} \| \Rho_1 - \Rho_2 \|_\Trace \ 
 = \ \frac{1}{2} \Trace \left| \Rho_1 -  \Rho_2 \right|
\end{align}
where the modulus for a selfadjoint operator with spectral decomposition $\bm{A} = \sum_i a_i \ketbra{i}{i}$ reads \cite{Reed-Simon_1972,Nielsen-Chuang_2000}
\begin{align}
 \left| \bm{A} \right| \ = \ \sqrt{\bm{A}^\dagger \bm{A}} \ = \ \sum_i |a_i| \  \ketbra{i}{i}.
\end{align}
We remark that other measures for distinguishability have also been investigated, such as the Helstrom matrix \cite{Chruscinski2011,Wissmann2015} and entropic measures \cite{Settimo2022,Megier2021}. We also note that a general set of conditions which have to be met by measures of distinguishability to quantify information flow can be found in Ref.~\cite{Smirne2022}. 

The trace distance has a direct physical meaning as a measure for distinguishability in the following sense. Suppose one party, Alice, has prepared the open system in either state $\Rho_1$ or $\Rho_2$ and sends it to a second party, Bob, who performs a measurement on the system in order to find the state prepared by Alice. It turns out that the maximal success probability for Bob to identify the correct state by an optimal strategy is given by
\begin{align} \label{p_opt}
 p_\mathrm{opt} = \ \frac{1}{2} \br{1 + \dis_\Trace\br{\Rho_1, \Rho_2}}.
\end{align}
The trace distance is bounded from below by $0$ (iff both state are the same) and from above by $1$ (iff both states are orthogonal). Thus, Eq.~\eqref{p_opt} shows that for orthogonal states Bob can identify the state with certainty. Under unitary evolutions the trace distance remains constant, reflecting that without coupling to the environment no information can leak out of the system. Finally, under completely positive and trace preserving maps (quantum channels) the trace distance can never increase, showing that a noisy quantum channel in general reduces the distinguishability of quantum states.

\begin{figure}[htp]
\centering
\includegraphics[width=0.9\columnwidth]{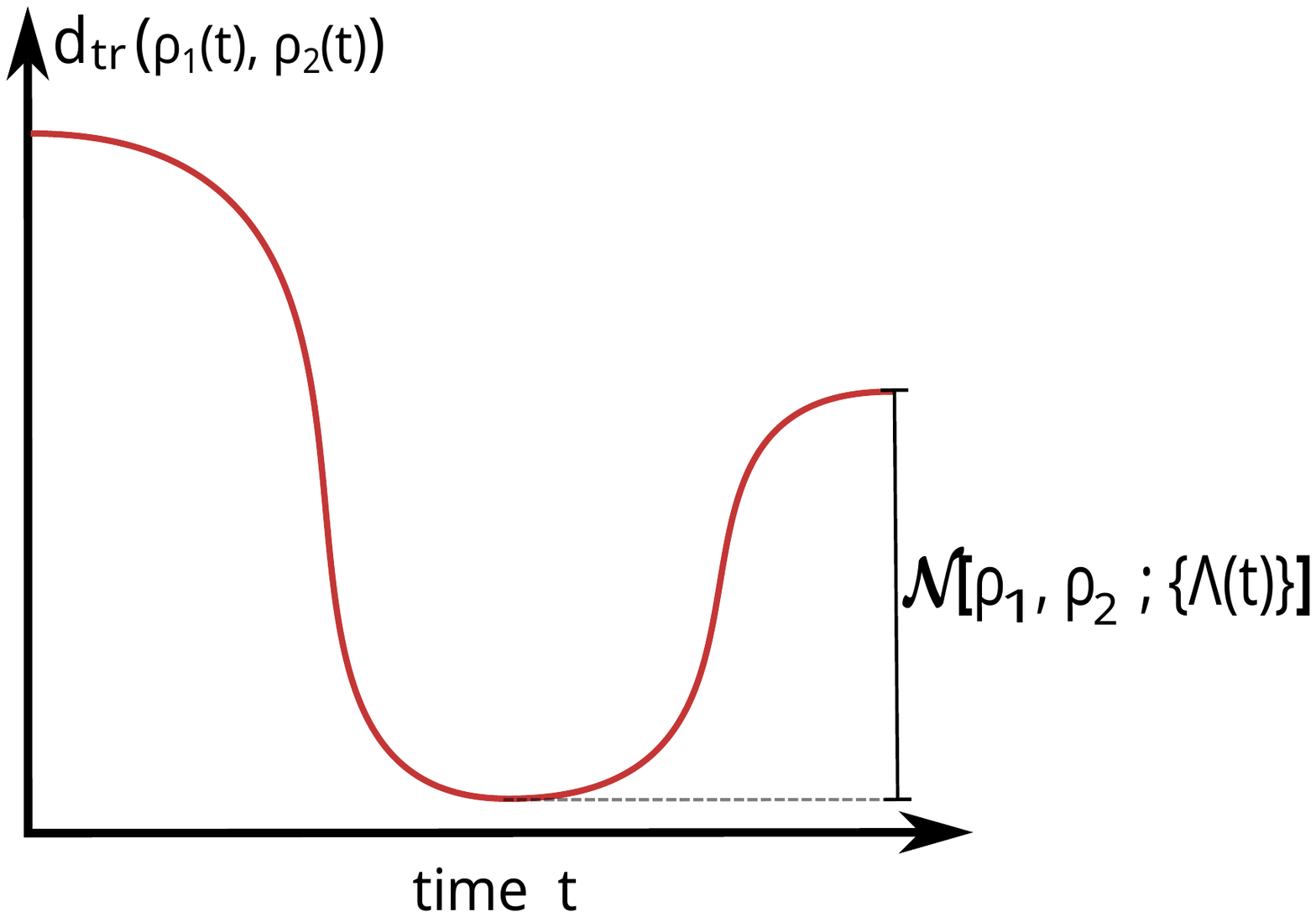}
\caption{\label{fig:Figure-1} Visualization of Eq.~(\ref{eq:N_measure_rho}) defining a measure of information flow from the environment back to the system.}
\end{figure}

Employing the trace distance as suitable quantity of distinguishability one can now define measures for the amount of information backflow as \cite{Breuer2009}
 	\begin{align}
 	\begin{split}\label{eq:N_measure_rho}
	&\mathcal{N} [\Rho_1, \Rho_2; \{\Lambda(t)\}] \ = \  \underset{\sigma \geq 0}{\int} dt \ \sigma(\Rho_1, \Rho_2, t)
	\\
	&\with \quad \sigma(\rho_1, \rho_2, t) \ := \ \frac{d}{dt} \dis_\Trace \br{ \Lambda(t) \Br{\Rho_1}, \Lambda(t) \Br{\Rho_2} }.
	\end{split}
	\\
	&\und \quad \tilde{\mathcal{N}} [\{\Lambda(t)\}]  \ = \ \underset{\rho_1, \rho_2}{\mathrm{max}} \ \mathcal{N} [\Rho_1, \Rho_2; \{\Lambda(t)\}]
	\label{eq:N_measure}
	\end{align}
The quantity in Eq.~(\ref{eq:N_measure_rho}) sums up the differences between minima and subsequent maxima for a pair of initial states $\Rho_1$ and $\Rho_2$ as illustrated in Fig.~\ref{fig:Figure-1}.
The maximization over the two initial states in Eq.~(\ref{eq:N_measure}) removes the dependence of the measures from them and indicates the maximal amount of information backflow the dynamical maps $\{\Lambda(t)\}$ can cause. To find those optimal initial state pairs can in general be difficult but they must be orthogonal to each other and need to be at the boundary of the convex set of states \cite{Wissmann2012}. This measure or similar ones found already several applications until recently, like to the Caldeira-Leggett model of quantum Brownian motion (which we will also address in Sec.~\ref{Application to the Caldeira-Leggett model}) \cite{Einsiedler2020} or the spin-boson model \cite{Clos2012,Wenderoth2021} but is also used in the context of quantum collision models \cite{Campbell2021} or the self-discharge of quantum batteries \cite{Xu2024}. Many further examples are discussed in Ref.~\cite{Breuer2016}.

\subsection{Classical measures of information flow}
\label{A Classical Measures of Information Flow}
Looking again on the notion of information flow as given in Sec.~\ref{Open quantum systems}, we find that its central idea actually has nothing particular quantum to it. For classical systems coupled to a bath or general environment any change of distinguishability can also be interpreted as loss or gain of information. Thus, it is natural to look for a suitable measure of distinguishability for the classical system. In this section we want to find such a measure, i.e. a suitable distance measure for classical phase space representations as used in the context of Hamiltonian mechanics. Thus, we describe classical states by probability distributions $W(\x)$ of phase space coordinates $\x = (q,p)^T \in \Gamma$, representing position and momentum. Hence, we basically have to deal with the time evolution of distributions on phase space $\Gamma$ and how to define information flow for them, as already done in \cite{Smirne2013}. We summarize it here and apply it directly to the specific case of phase space. Instead of orthogonal support as for density operators, perfectly distinguishable states would be given by distributions $W_{1,2}$ with disjunct support. Thus, for such two $W_{1,2}$ the distance should again be one. Furthermore, any canonical transformation in phase space should not change anything about the distinguishability analogously to unitary maps in the quantum case, i.e. the distance should be constant under canonical transformations. Finally, so called stochastic maps on phase space, i.e. convex linear maps conserving the properties of probability distributions, should be contracting under a suitable distance equivalently to CPTP maps on quantum systems contracting under the trace distance. A distance fulfilling all those requirements is given by the \textit{Kolmogorov distance} \cite{Smirne2013}
	\begin{align}\label{eq:Koldis}
	\dis_\Kol (W_1, W_2) \ = \ \frac{1}{2}\underset{\Gamma}{\int} dx \ \left| W_1(\x) - W_2(\x) \right|.
	\end{align}
where the invariance under canonical transformations follows from Liouville's theorem.

We remark that for discrete probability distributions written as probability vectors $\vec{v}$ and $\vec{w}$  with components $v_i$ and $w_i$ we have
	\begin{align}
	\dis_\Kol (\vec{w}, \vec{v}) \ = \ \frac{1}{2}\sum_i \ \left| w_i - v_i \right|.
	\end{align}
We further note that this Kolmogorov distance is closely connected to the trace distance. Since quantum measurements can be defined as \textit{positive operator valued measures} (POVM) $\{ \E_i \}$ where $\E_i$ is the so called \textit{effect operator} representing the measurement of outcome $i$ with probability $p_i = \Tr{\E_i^\dagger \E_i \Rho}$ depending on the state of the system, one can show that \cite{Nielsen-Chuang_2000}
	\begin{align}\label{eq:dtr_by_dkol}
	\dis_\Trace (\Rho_1, \Rho_2) \ = \ \underset{\{ \E_i \}}{\max} \ \dis_\Kol (\vec{p}_1, \vec{p}_2),
	\end{align}
where the maximum is to be understood as the maximum over all possible quantum measurements (i.e POVMs) and $\vec{p}_i$ is the probability vector of $\{ \E_i \}$ when the system is prepared in state $\Rho$.

\section{Continuous Variable Quantum Systems}
\label{Continuous Variable Quantum Systems}
In this section we consider the class of quantum systems whose states can be represented by quasi-probability distributions on phase space, and how the two measures of information flow \--- the quantum one based on the trace distance and the classical one based on the Kolmogorov distance \--- are connected. These so-called continuous variable quantum (CVQ) systems \cite{Adesso2014} are defined on an infinite dimensional Hilbert space equipped with quadrature operators, i.e. generalized position and momentum operator, with canonical commutation relation
	\begin{align}
	[\bm{q}, \bm{p}] = i \hbar \ \1.
	\end{align}
To get the phase space representation of a CVQ system one needs two additional operators, called the \textit{s-ordered Weyl-displacement operator} $\bm{\mathcal{D}}^s(\x)$ and its Fourier transform $\bm{\mathcal{T}}^s(\x)$ defined by \cite{Cahill1969a, Cahill1969b}
	\begin{align}
	\bm{\mathcal{D}}^s(\x) \ := \ \exp \left[ -\frac{i}{\hbar} (\Omegabm\x) ^T  \X + \frac{s}{4\hbar}\| \x \|^2 \right],
	\label{eq:D(x)}
	\\
	\bm{\mathcal{T}}^s(\x) \ := \ \frac{1}{(2 \pi \hbar)^2} \underset{\Gamma}{\int} d\vec{y} \ \bm{\mathcal{D}}^s(\vec{y})  \exp \left[ -\frac{i}{\hbar} (\Omegabm\y)^T \x \right],
	\label{eq:T(x)}
	\end{align}
where $\Omegabm := \left(\begin{array}{cc} 0 & 1 \\ -1 & 0 \end{array}\right)$ is the symplectic form of the phase space, $\X := (\q, \p)^T$ simply contains position and momentum operator and the order parameter $s$ takes values in the interval $[-1, 1]$. Thus, one can represent a quantum state $\Rho$ by its $s$-ordered quasi-probability distributions on phase space defined by
	\begin{align}\label{eq:W^s_by_rho}
	W_\rho^s(\x) \ := \ \Tr{\bm{\mathcal{T}}^s(\x) \Rho},
	\end{align}
which reversely can be used to decompose a quantum state
	\begin{align}\label{eq:rho_by_W^s}
	\Rho \ := \ \int_\Gamma d\x \ W_\rho^s(\x) \bm{\mathcal{T}}^{-s}(\x),
	\end{align}
as is explained in detail in Ref.~\cite{Cahill1969a}. This means that
the operators $\{\bm{\mathcal{T}}^{s}(\x)\}$ form an over-complete and non-orthogonal basis -- or \textit{frame} \cite{Kovacevic2008} -- with the so-called dual basis or \textit{dual frame} $\{\bm{\mathcal{T}}^{-s}(\x)\}$.

The expression "quasi"-probability distributions is used because they are real and integrate to one, yet might be locally negative and since different points in phase space do not represent, from a quantum measurement point of view, mutually excluding events. For any ordering parameter $s$ these distributions are unique representations of the state $\Rho$, though usually only three of them are commonly used: For $s=-1$ one has the \textit{Husimi Q-function} \cite{Husimi1940}, for $s=0$ the \textit{Wigner function} \cite{Wigner1932} and finally for $s=1$ one gets the \textit{Glauber P-function} \cite{Glauber1963,Sudarshan1963} (for a detailed exposition, see Ref.~\cite{Cahill1969b}).

Furthermore, we remark that in context of Husimi Q- and Glauber P-functions one usually represents points in phase 
space by complex numbers where the real part is associated with the position coordinate and the imaginary part with the momentum, a representation used and well explained e.g. in Ref.~ \cite{Adesso2014}. In the present paper, however, we represent the phase space as a two-dimensional real vector space. Additionally there exist several conventions about the value of $\hbar$; for example Ref.~\cite{Adesso2014} actually uses $\hbar = 2$ instead of the more usual $\hbar = 1$, something one should keep in mind when comparing pre-factors. In this paper we decided to keep $\hbar$ as an explicit constant and in some sense as a natural unit of phase space volume. In appendix \ref{Representing phase space in complex plane and position-momentum coordinates} we summarize the translation from the position-momentum coordinate representation to the complex plain representation to ease comparison with other publications concerning CVQ systems.

A prominent class of states for CVQ quantum systems are so called \textit{Gaussian quantum states}, i.e. states with Gaussian shaped Wigner function $W = \Gauss{\vecmu}{\sigmabm}$ completely characterized by their first statistical moments encoded in the \textit{displacement vectors} $\vecmu := \braket{\X}$ and second statistical moments encoded in the \textit{covariance matrix} $\sigmabm_{ij} := \braket{X_i X_j + X_j X_i} - 2\braket{X_i}\braket{X_j}$.

\section{Quantum-to-Classical Transition} \label{Quantum-to-Classical Transition}
Having a class of quantum systems which \--- in addition to their Hilbert space representation \--- can also be described by a phase space distribution, the question arises about any connection between the two introduced measures of information flow, the trace distance based quantum measure and the Kolmogorov distance based classical one. We already addressed this question in a previous paper \cite{Richter2022} and showed that the trace distance between two density operators $\Rho_1$ and $\Rho_2$ is bounded by the Kolmogorov distances between their corresponding Husimi Q- and Glauber P-functions $W_i^{-1} =: Q_i$ and $W_i^{+1} =: P_i$ as
	\begin{align}
	\dis_\Kol(Q_1, Q_2) \ \leq \ \dis_\Trace(\Rho_1, \Rho_2) \ \leq \ \dis_\Kol(P_1, P_2).
	\label{eq:tracedis_sandwich}
	\end{align}
We used this chain of inequalities already to define a minimal approximation of the trace distance induced measure of information backflow based on Husimi Q- and Glauber P-functions \cite{Richter2022}. We will now go even further and state that for two quantum states $\Rho_1$ and $\Rho_2$ both represented by quasi-probability distributions of order $r$ and $s$ with $r<s$ the respective Kolmogorov distances fulfill
	\begin{align}\label{eq:d(s)<d(r)}
	\dis_\Kol(W_1^r, W_2^r) \ \leq \ \dis_\Kol(W_1^s, W_2^s),
	\end{align}
i.e. the higher the order parameter the larger the respective Kolmogorov distance. For the proof and the remaining text we set the following notation of a two dimensional Gaussian distribution $\Gauss{\vecmu}{\sigmabm}$ with displacement vector $\vecmu$ and covariance matrix $\sigmabm$
\begin{align}
\Gauss{\vecmu}{\sigmabm} (\x) \ := \ \frac{\exp \Br{ -\br{\x - \vecmu}^T \sigmabm^{-1} \br{\x - \vecmu} }}{\pi \ \sqrt{\det \sigmabm}}
\label{eq:2D_Gaussian}
\end{align}
with missing displacement denoting $\vecmu = 0$ and missing covariance matrix $\sigmabm = \hbar \1$ as default.

\begin{proof}
To cut the proof to the essential let us first just state the fact that for any state the differently ordered quasi-probability distributions can be transformed from one to the other by convolution with a Gaussian kernel, i.e. for $r < s$ we have
	\begin{align}\label{eq:r->s_convolution}
	\begin{split}
	W^r(\x) \ &= \ \int d\x \ W^s(\vec{y}) \cdot \frac{\exp \left[-\frac{\|\x - \vec{y}\|^2}{(s-r)\hbar} \right]}{\pi \ (s-r)\hbar}
	\\
	&=: \ \left(W^s \star \Gauss{}{(s-r)\hbar \1} \right) (\x)
	\end{split}
	\end{align}
with $\star$ denoting the convolution and
\begin{align}
\Gauss{}{(s-r)\hbar \1} (\x) \ := \ \frac{\exp \left[-\frac{\|\x\|^2}{(s-r)\hbar} \right]}{\pi \ (s-r)\hbar}.
\end{align}
in accordance to Eq.~(\ref{eq:2D_Gaussian}). The proof for this can be found in appendix \ref{Transformation between differently ordered quasi-probability distributions}. The proof of Eq.~(\ref{eq:d(s)<d(r)}) is now  easily obtained as follows:
	\begin{align}
	\begin{split}
	&\dis_\Kol(W_1^r, W_2^r) \ = \ \int d\x \ \left| W_1^r(\x) - W_2^r(\x) \right|
	\\
	= \ &\int d\x \ \left| \int d\vec{y} \ubs{\geq 0}{\Gauss{}{(s-r)\hbar \1}(\x - \vec{y})} \left( W_1^s(\vec{y}) - W_2^s(\vec{y}) \right) \right| 
	\\
	\leq \ &\int d\vec{y} \ubs{=1 \ \forall y}{\int d\x \ \Gauss{}{(s-r)\hbar \1}(\x - \vec{y})} \left| W_1^s(\vec{y}) - W_2^s(\vec{y}) \right| 
	\\
	= \ &\int d\vec{y} \ \left| W_1^s(\vec{y}) - W_2^s(\vec{y}) \right| \ = \ \dis_\Kol(W_1^s, W_2^s).
	\end{split}
	\end{align}
\end{proof}

As a consequence of Eq.~(\ref{eq:tracedis_sandwich}) and Eq.~(\ref{eq:d(s)<d(r)}) and due to the continuous transformation from one ordering to the other, for any pair of states $\rho_1$ and $\rho_2$ in a continuous variable quantum system \--- represented either by their density operators $\Rho_{1,2}$ or their s-ordered quasi-probability distributions $W^s_{1,2}$ \---  there exists a unique optimal ordering $\sopt \in [-1,1]$ such that
	\begin{align}\label{eq:optimal_ordering}
	\dis_\Trace(\Rho_1, \Rho_2) \ = \ \dis_\Kol(W_1^{\sopt} , W_2^{\sopt}).
	\end{align}
Thus,  there is always a unique optimal ordering parameter for which the Kolmogorov distance of the corresponding quasi-probability distributions coincides with the trace distance of the density matrices.

\begin{figure}[htp]
\centering
\includegraphics[width=\columnwidth]{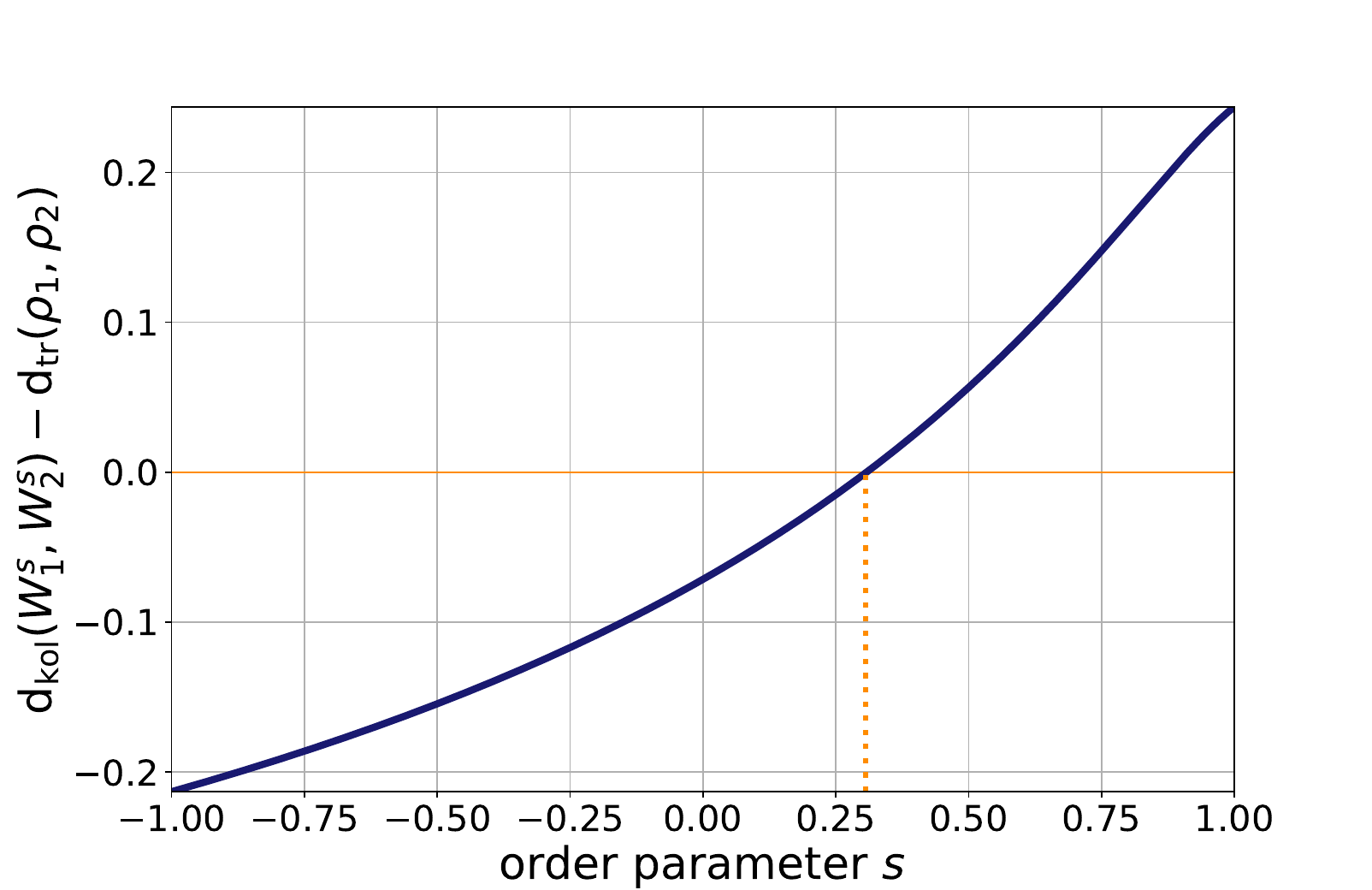}
\caption{ Deviation from the trace distance of the Kolmogorov distance of the $s$-ordered quasi-probability distributions over the ordering parameter $s$, such that the intersection with the $s$-axis marks the optimal ordering according to Eq.~(\ref{eq:optimal_ordering}) The two states are Gaussian states with centered around $\x_{1,2} = \pm (0.75, \ 0.0)^T$ and Wigner function covariance matrix $\sigmabm_{1,2} = 1.2\hbar \cdot \1$.}
\label{fig:Figure-2}
\end{figure}

Figure \ref{fig:Figure-2} illustrates Eq.~(\ref{eq:optimal_ordering}) which shows for an exemplary pair of Gaussian quantum states $\dis_\Kol (W_1^s, W_2^s) - \dis_\Trace (\Rho_1, \Rho_2)$ over the ordering parameter $s$ such that the intersection of the graph with the $s$-axis marks the optimal ordering $\sopt$. To compute the density operator of a Gaussian state given its displacement vector and covariance matrix we used the python package The Walrus \cite{TheWalrus}.

In the following we want to discuss the transition of the quantum measure to the classical phase space measure. To do so we will use the following definitions.

\begin{definitions*}
\item[i)]
A \ul{Gaussian-mixed state} is a state $\Rho$ of a continuous variable quantum system which can be decomposed into a convex combination of Gaussian states
	\begin{align}\label{eq:Gaussian-mixed-state}
	\Rho \ = \ \sum_i \ p_i \cdot \G{\vecmu_i}{\sigmabm_i}
	\end{align}
with $(p_i)$ being a discrete probability distribution. It is thus completely characterized by $(p_i)$ and the statistical moments $(\vecmu_i)$ and $(\sigmabm_i)$.
\item[ii)]
A state $\Rho$  is said to be a \ul{state of classical uncertainty} iff it is a Gaussian-mixed state defined by Eq.~(\ref{eq:Gaussian-mixed-state}) with $\sigmabm_i \gg \hbar \1 \ \forall i$\footnote{The expression $\sigmabm \gg \hbar \1$ means that all eigenvalues of $\sigmabm$ are much larger than $\hbar$.}.
\item[iii)]
Accordingly, we define the \ul{limit of classical uncertainty} for  Gaussian-mixed states, denoted as
\begin{align}
\sigmabm(\Rho)/\hbar \to \infty,
\end{align}
by demanding that all covariance matrices $\sigmabm \in \{\sigmabm_i\}$ in Eq.~(\ref{eq:Gaussian-mixed-state}) fulfill the limit
	\begin{align}
	\frac{\sigma}{\hbar} \to\infty \  \forall \sigma \in \mathrm{spec}[\sigmabm].
	\end{align}
\label{def:classical-uncertain}
\end{definitions*}

Due to Eq.~(\ref{eq:W^s_by_rho}) the quasi-probability distributions for Gaussian-mixed states take the form
	\begin{align}
	W_\rho^s (\x) \ = \ \sum_i \ p_i \cdot \Gauss{\vecmu_i}{\sigmabm_i - s\hbar \1}.
	\end{align}
Using the definitions above we find that for Gaussian-mixed states $\Rho$ and for any two order parameters $r,s \in [-1,1]$ one has
	\begin{align}\label{eq:limit_Wr-Ws}
	W_\rho^s - W_\rho^r \ \xrightarrow{\sigmabm(\Rho)/\hbar \to \infty} \ 0,
	\end{align}
i.e. for states of classical uncertainty the quasi-probability distributions look the same no matter the ordering.

\begin{proof}
It will be enough to show that the Glauber P-function converges for states of classical uncertainty to the Husimi Q-functions since those two functions with $s = \pm1$ have the two extreme orderings and the transitions between one ordering to the other is a continuous and monotonic transition. This can be seen directly from the convolution formula in Eq.~(\ref{eq:r->s_convolution}) stating that the transition from ordering $s$ to $r <s$ is just a convolution with a Gaussian of width $(s-r)\hbar$ and will monotonically widen the distribution.

For the Husimi Q-function we thus find
	\begin{align}
	Q_\rho \ = \ P_\rho \star \Gauss{}{2 \hbar \1} \ = \ \sum_i \ p_i \cdot \Gauss{\vecmu_i}{\sigmabm_i - \hbar \1} 	\star \Gauss{}{2 \hbar \1}
	\end{align}
By definition we know that for all $i$ the eigenvalues of the covariance matrices $\sigmabm_i$ are all much larger than $\hbar$ and, thus, we first have $\sigmabm_i - \hbar \1 \simeq \sigmabm_i$. Secondly, all Gaussians $\Gauss{\vecmu_i}{\sigmabm_i}$ are much broader than the convolution kernel $\Gauss{}{2 \hbar \1}$ which in comparison is almost a delta peak and will hence not change the distribution convoluted with. Consequently we find
	\begin{align}
	Q_\rho \ = \ P_\rho \star \Gauss{}{2 \hbar \1} \ \simeq \ P_\rho,
	\end{align}
which proves the statement.
\end{proof}

As a consequence of Eq.~(\ref{eq:tracedis_sandwich}) and Eq.~(\ref{eq:limit_Wr-Ws}) we now find that for any pair of Gaussian-mixed states $\Rho_1$ and $\Rho_2$ and for any ordering $s \in [-1,1]$ in the limit of classical uncertainty we have
	\begin{align}\label{eq:limit_dkol}
	\dis_\Kol (W_1^s, W_2^s) \ \xrightarrow{\sigmabm(\Rho_{1,2})/\hbar \to \infty} \ \dis_\Trace(\Rho_1, \Rho_2).
	\end{align}
Thus, if one is concerned with quantum states whose uncertainty is significantly larger than the quantum limit $\hbar$ we can replace the trace distance of the density operator by the Kolmogorov distance of the their quasi-probability distributions of any ordering. In case of such initial states and some dynamics which will not drastically decrease their uncertainty we can accordingly measure the flow of information between the CVQ system and its environment in a very good approximation by the change of those Kolmogorov distances allowing to quantify the information flow purely by means of a representation in phase space. Moreover, we see how in the limit of classical uncertainty the quantum measure of information flow based on the trace distance between density operators converges to the measure within classical phase space allowing to compare the information flow resulting from dynamics of CVQ systems and the information flow from dynamics in classical phase space.

\section{Application to the Caldeira-Leggett model}
\label{Application to the Caldeira-Leggett model}
The application of the limit in Eq.~(\ref{eq:limit_dkol}) to information flow can be illustrated with the help of the Caldeira-Leggett model of quantum Brownian motion \cite{Caldeira1983,Grabert1988}. Here, the 
system-bath Hamiltonian is given by
	\begin{align}
	\begin{split}
	\bm{H}_\total \ &= \ \ubs{\bm{H}_\sys}{\frac{\p_0^2}{2m_0} + \frac{1}{2}m_0 \omega_0^2 \q_0^2} \ + \ \ubs{\bm{H}_\bath}{\sum_i \frac{\p_i^2}{2m_i} + \frac{1}{2}m_i \omega_i^2 \q_i^2}
	\\
	&- \ \ubs{\bm{H}_\inter}{\q_0 \otimes \sum_i \lambda_i \q_i} \ + \ \ubs{counter term}{\bm{q}^2_0  \sum_i \frac{\lambda_i^2}{2 m_i \omega_i^2}},
	\end{split}
	\end{align}
consisting of a central harmonic oscillator $\bm{H}_\sys$ linearly coupled to a bath of harmonic oscillators $\bm{H}_\bath$ via the position operators through $\bm{H}_\inter$, and a counter term to renormalize the system frequency. This model is a well-known standard example of open system dynamics of both quantum and classical systems. The Hamiltonian is quadratic and preserves Gaussianity of states, which enables a simple analytical solution \cite{Breuer-Petruccione_2007}. Nevertheless, the model shows many interesting features of open systems, in particular non-Markovian quantum dynamics \cite{Einsiedler2020,Ma-Arb_Einsiedler_2020}. Note that in Sec. \ref{Continuous Variable Quantum Systems} and Sec. \ref{Quantum-to-Classical Transition} mass and frequency where not written explicitly, giving equal units of $\sqrt{\hbar}$ to position and momentum operators. In this section we write mass $m_0$ and frequency $\omega_0$ explicitly, with the corresponding transformation is given by
\begin{align}
\q \ = \sqrt{m_o \omega_0} \q_0 \quad \und \quad \p = \frac{1}{\sqrt{m_o \omega_0}} \p_0.
\end{align}

Assuming a continuous bath with Ohmic spectral density and Lorentz-Drude cutoff
	\begin{align}
	J(\omega) \ = \ \frac{2m_0 \gamma}{\pi} \omega \frac{\Omega^2}{\Omega^2 + \omega^2},
	\end{align}
where $\gamma$ describes the coupling strength between system and bath and $\Omega$ the frequency cutoff, the model can be solved analytically. We followed the solutions as used for example in \cite{Einsiedler2020} and presented in more detail in \cite{Breuer-Petruccione_2007,Ma-Arb_Einsiedler_2020}. To compute the density operators of Gaussian states we again used the package The Walrus \cite{TheWalrus}.

\begin{figure*}[htp]
\subfloat{\includegraphics[width=0.5\linewidth]{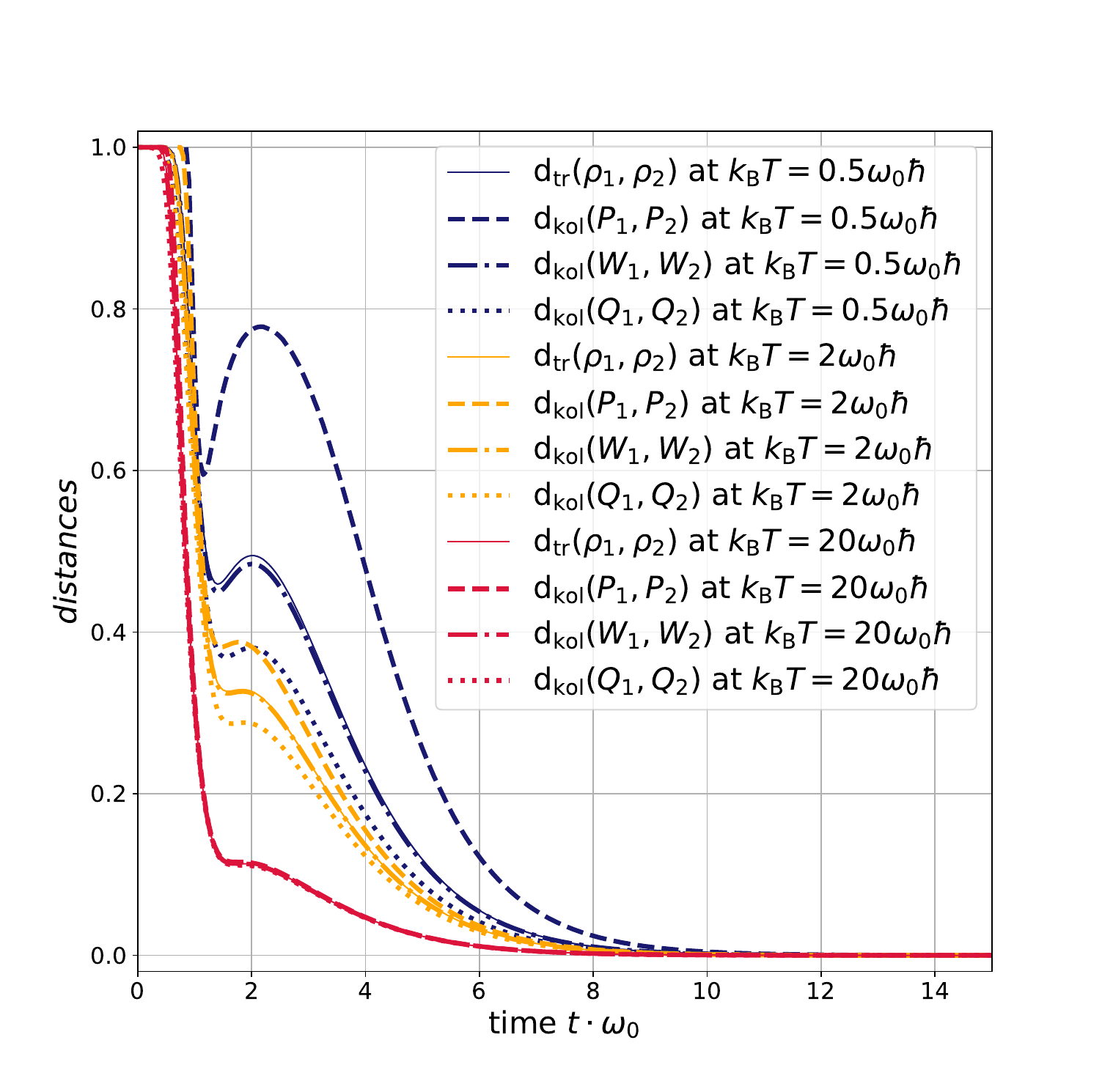}}\hfill
\subfloat{\includegraphics[width=0.5\linewidth]{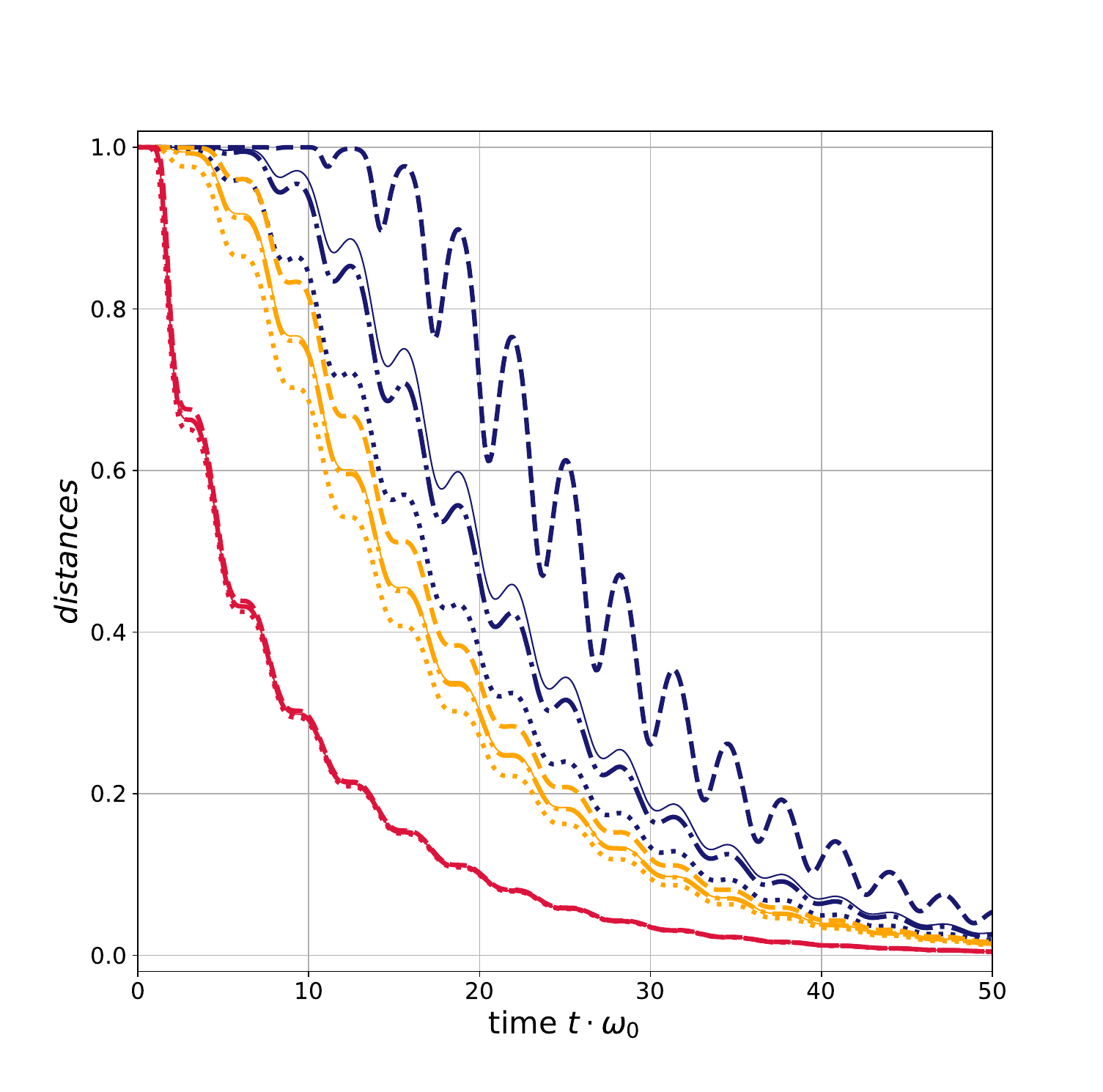}}\\[1em]
\subfloat{\includegraphics[width=0.5\linewidth]{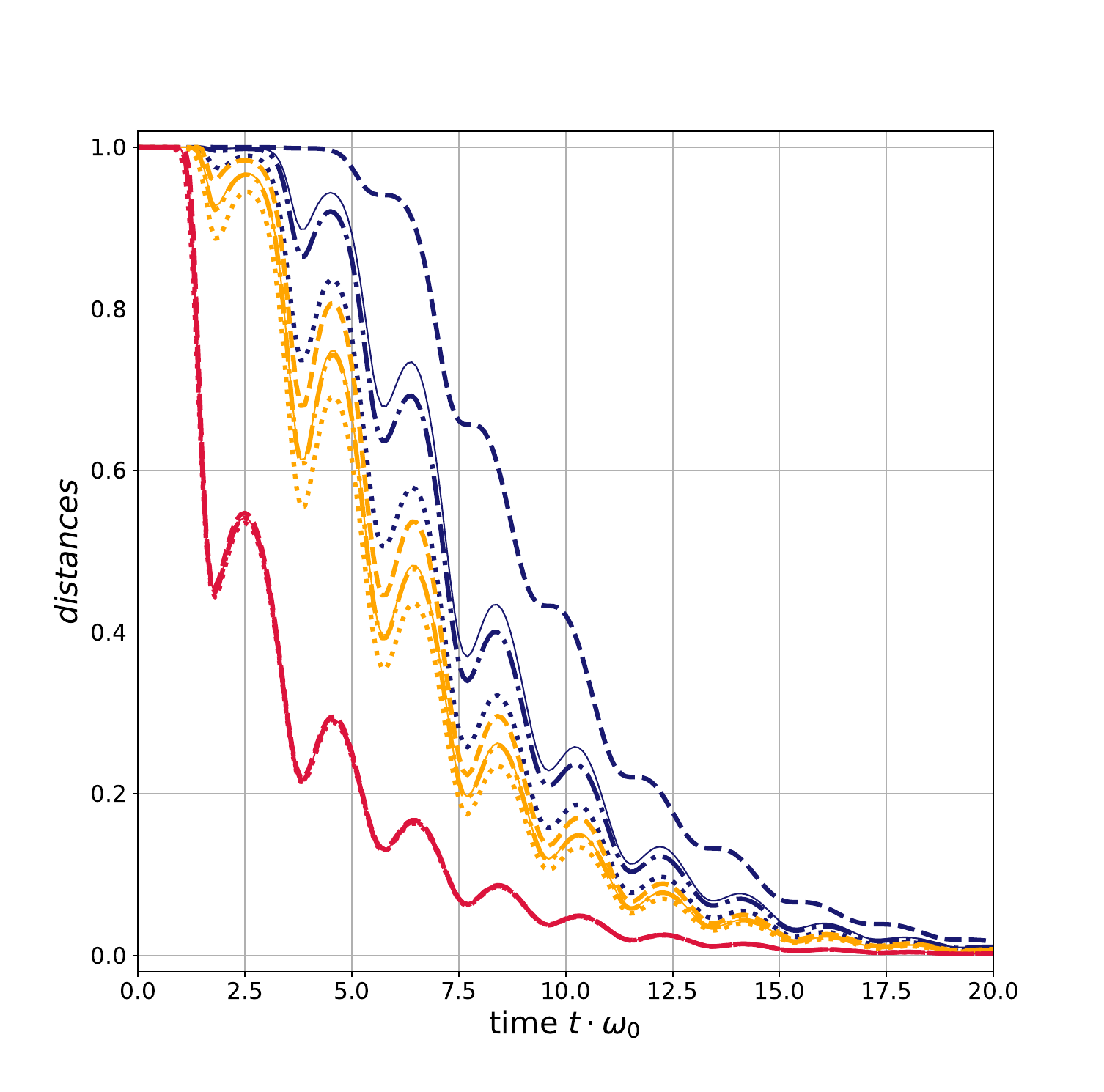}}\hfill
\subfloat{\includegraphics[width=0.5\linewidth]{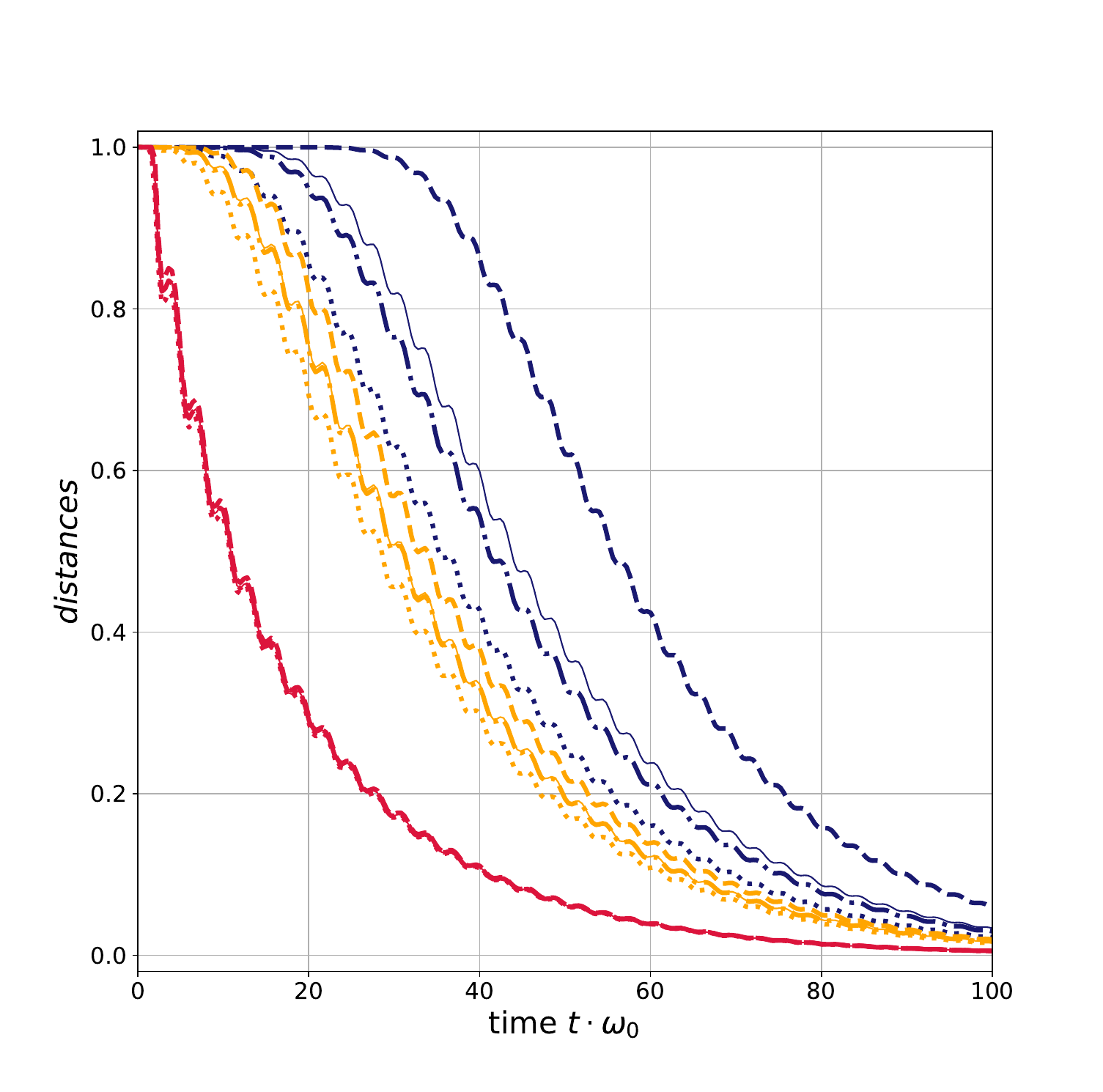}}
\caption{Trace distance and Kolmogorov distances of P, W and Q functions over time for two initially coherent states $\ketbra{\pm\x}{\pm\x}$ with $\x = (4.0/\sqrt{2m_0\omega_0}, \ 0.0)^T$ for different temperatures. Upper left: scaling limit $\Omega = 100 \omega_0$ and strong coupling $\gamma = \omega_0$; upper right: scaling limit $\Omega = 100 \omega_0$ and weak coupling $\gamma = 0.1 \omega_0$;  lower left: low cutoff $\Omega = \omega_0$ and strong coupling $\gamma = \omega_0$; lower right: low cutoff $\Omega = \omega_0$ and weak coupling $\gamma = 0.1 \omega_0$.}
\label{fig:dis-CL}
\end{figure*}

\begin{figure*}[htp]
\subfloat{\includegraphics[width=0.5\linewidth]{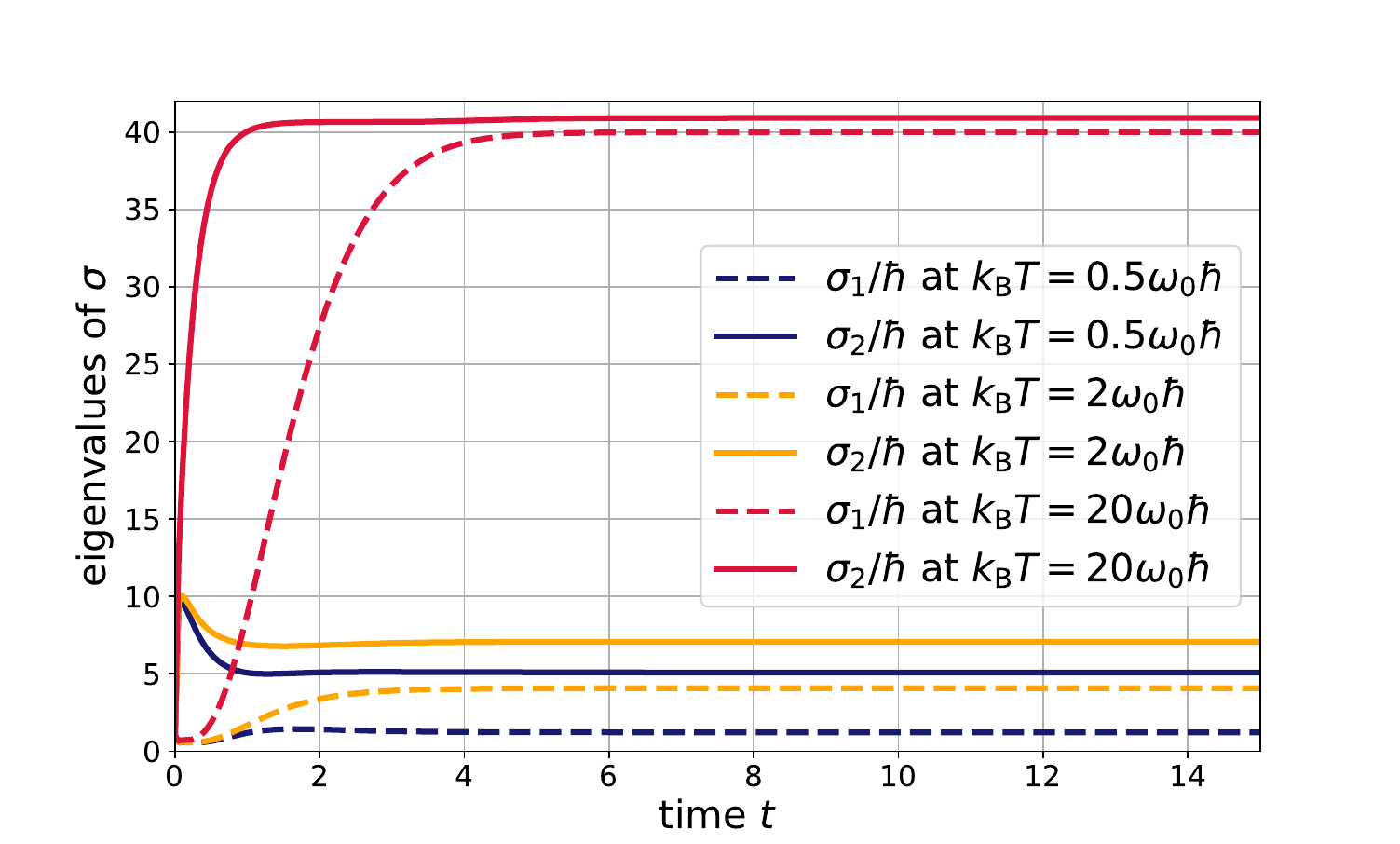}}\hfill
\subfloat{\includegraphics[width=0.5\linewidth]{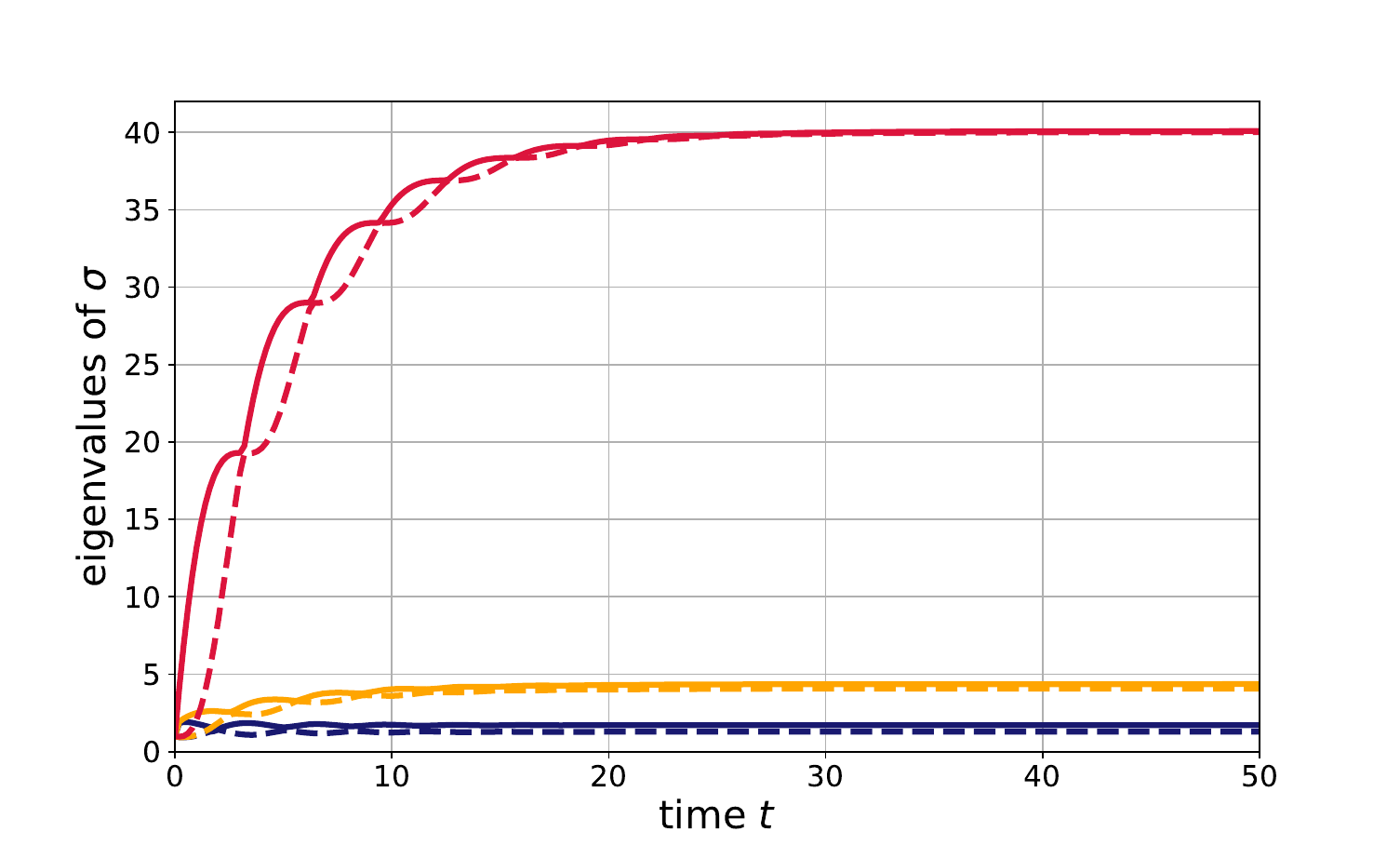}}\\[1em]
\subfloat{\includegraphics[width=0.5\linewidth]{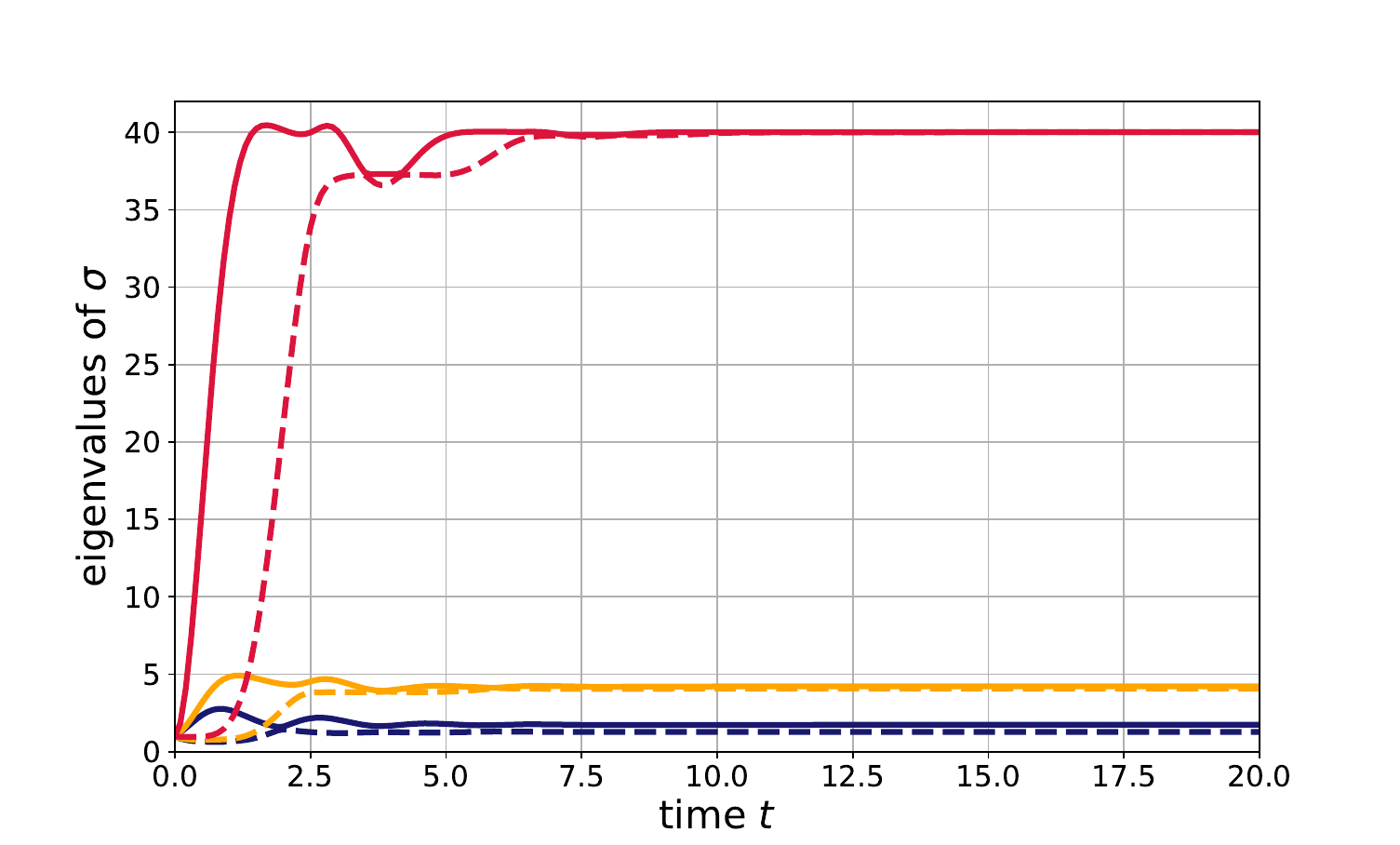}}\hfill
\subfloat{\includegraphics[width=0.5\linewidth]{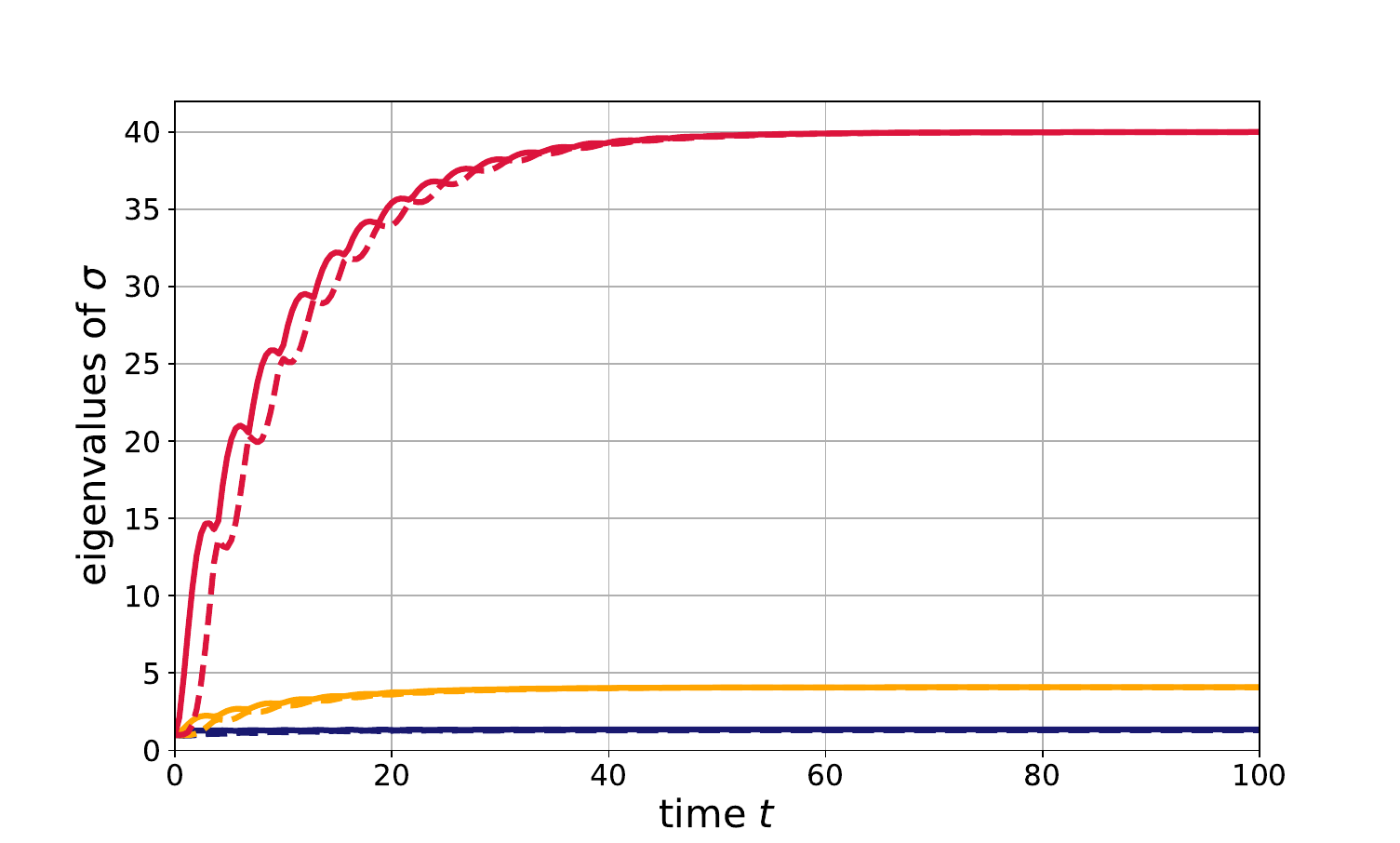}}
\caption{Eigenvalues of the covariance matrix over time for an initially coherent states (i.e. $\sigmabm = \hbar \1$ at time $t=0$) for different temperatures. Top left: scaling limit $\Omega = 100 \omega_0$ and strong coupling $\gamma = \omega_0$; top right: scaling limit $\Omega = 100 \omega_0$ and weak coupling $\gamma = 0.1 \omega_0$;  bottom left: low cutoff $\Omega = \omega_0$ and strong coupling $\gamma = \omega_0$; bottom right: low cutoff $\Omega = \omega_0$ and weak coupling $\gamma = 0.1 \omega_0$.}
\label{fig:eigenvalues_of_sigma}
\end{figure*}

In Fig.~\ref{fig:dis-CL} we plot as a function of time the Kolmogorov distances of the P-functions, the Wigner functions and the Q-functions, as well as the trace distance for two initially coherent states for three different temperatures. Additionally, we varied coupling strength $\gamma$ and frequancy cutoff $\Omega$ in different subplots: the upper left representing strong coupling in the scaling limit of high cutoff, the upper right weak coupling and high cutoff, the lower left strong coupling and small cutoff and the lower right weak coupling and small cutoff. In all four parameter regimes we see a strong dependency on the temperature of the difference between Kolmogorov distances and trace distance. For low temperatures ($\kbT = 0.5 \omega_0 \hbar$) especially the Kolmogorov distance for the P and the Q functions differ significantly from the trace distance while for high temperatures ($\kbT = 20 \omega_0 \hbar$) almost no gap between them is visible. This is due to the effect of the bath temperature on the covariance matrix $\sigmabm$ as depicted in Fig.~\ref{fig:eigenvalues_of_sigma} plotting its eigenvalues of $\sigma_1$ and $\sigma_2$ over time for the same temperatures, coupling strengths and frequency cutoffs as in Fig.~\ref{fig:dis-CL}.

The higher the temperature of the bath the larger are the eigenvalues of the covariance matrix in the stationary state. One also sees that in all four regimes of coupling strength and frequency cutoff the eigenvalues grow rapidly relative to the corresponding time range for each regime. Thus, for high temperatures one quickly reaches the limit of classical uncertainty where the difference between the quasi-probability distributions of different ordering becomes negligible according to Eq.~(\ref{eq:limit_Wr-Ws}).

A remarkable feature is the performance of the Kolmogorov distance of the Wigner functions (in the following called \textit{Wigner distance}) as approximation of the trace distance. In all four subplots we find that even in case of low temperature the Wigner distance deviates from the trace distance much less than the Kolmogorov distances of the P and Q functions and already in the case of $\kbT = 2 \omega_0\hbar$ it matches the trace distance almost perfectly in all four regimes, although the Kolmogorov distances of the P and Q functions still differ significantly from the trace distance. This suggests that for increasing temperature the optimal ordering of Eq.~(\ref{eq:optimal_ordering}) shifts towards zero, i.e. the ordering of the Wigner function. This is an interesting point which clearly deserves further investigations.

\begin{figure*}[htp]
\includegraphics[width=0.9\textwidth]{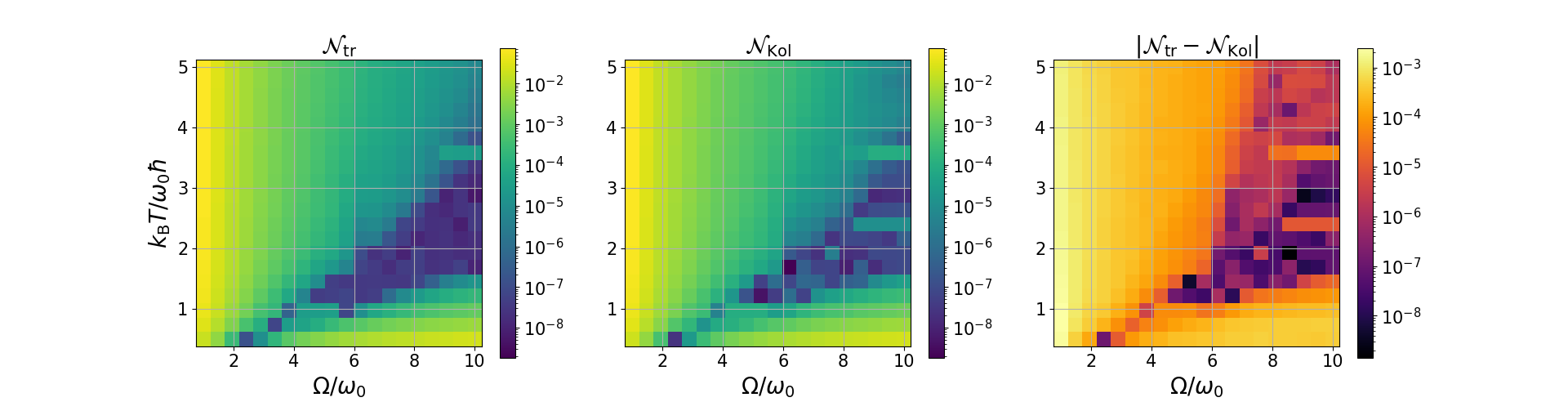}
\\
\includegraphics[width=0.9\textwidth]{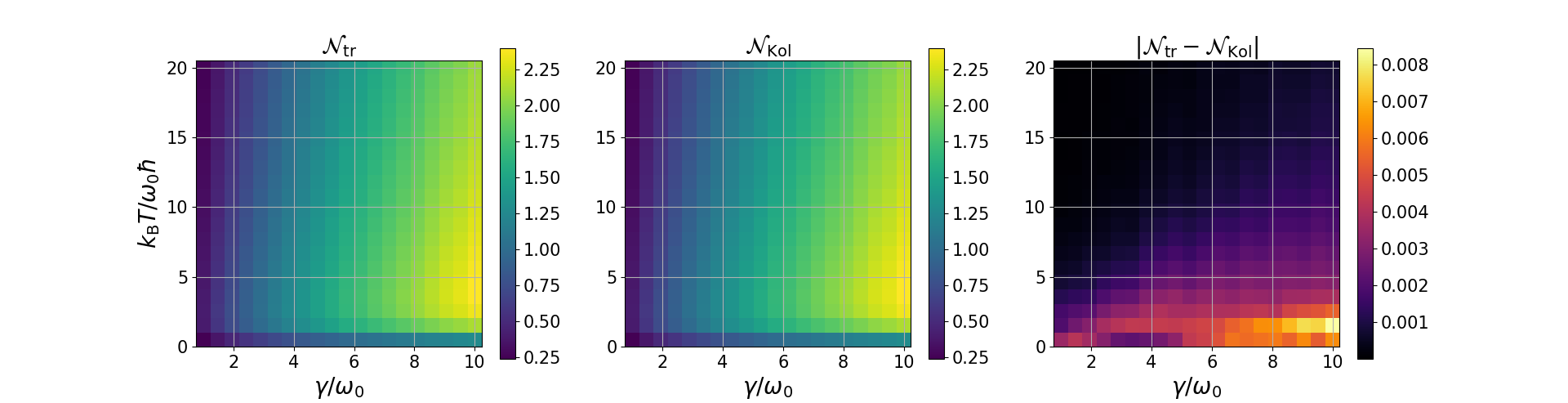}
\caption{Information backflow measured by the trace distance $\mathcal{N}_\mathrm{tr}$ (left), approximated using the Kolmogorov distance of the Wigner functions $\mathcal{N}_\mathrm{kol}$ (middle) and the absolute error $\vert \mathcal{N}_\mathrm{tr} -  \mathcal{N}_\mathrm{kol} \vert / \mathcal{N}_\mathrm{tr}$ of the approximation (right) over temperature $\kbT$ and frequency cutoff $\Omega$ at coupling strength $\gamma = 0.1 \omega_0$ (upper row) and over temperature $\kbT$ and coupling strength $\gamma$ at frequency cutoff $\Omega = \omega_0$ for two initially coherent states $\ketbra{\pm\x}{\pm\x}$ with $\x = (4.0/\sqrt{2m_0\omega_0}, \ 0.0)^T$. The time interval taken into account is $t \omega_0 \in [0.0, 50.0]$ in the upper row and $t \omega_0 \in [0.0, 250]$ in the lower row. Please note that the scales of the errors differ from the scales of $\mathcal{N}_\mathrm{tr}$ and $\mathcal{N}_\mathrm{kol}$ also signaled by the different color map.}
\label{fig:Test_Nkol_vs_Ntr}
\end{figure*}

The plots in Fig.~\ref{fig:dis-CL} propose the Wigner distance as a practical approximation of the trace distance in various settings of parameters, especially for higher temperatures. This is in accordance with the use of the Wigner function in semi-classical scenarios of quantum mechanics \cite{Hillery1984}. In the last part we will thus test the application of the Wigner distance to measures of information backflow on two interesting set of parameters, depicted in Fig.~\ref{fig:Test_Nkol_vs_Ntr}.  In the weak coupling regime Ref.~\cite{Einsiedler2020} showed when plotting the information backflow over temperature and frequency cutoff one encounters an area of minimal backflow. For the spin-boson model a similar effect was observed in Ref.~\cite{Clos2012} and explained by resonance effects of the system oscillator with the bath at an effective temperature depending spectral density. We plotted for this area the measure of Eq.~(\ref{eq:N_measure_rho}) for two initially coherent states in the upper row of Fig.~\ref{fig:Test_Nkol_vs_Ntr}, showing the trace distance (left), the approximation via the Wigner distance (middle) and finally the absolute error of the approximation (right).

One can see that the fundamental structure is well reflected by the approximation using the Wigner distance. Especially for settings causing larger amounts of information backflow the approximation is pretty close. However, the valley of low information backflow is reflected less clearly. We account this at least partially as a result of numerical errors both in the truncation of the density operator as well as the truncation of the integration area of the Wigner Kolmogorov distances. Both truncations limit the accuracy of the corresponding distances which are in general very low yet become visible on the logarithmic scale in this area of low information backflow. Nevertheless is the absolute error almost over the whole plot at least one decimal power smaller than the measure $\mathcal{N}_\mathrm{tr}$.

The second set of parameters on which we want to test the Wigner distance is shown in the lower row of Fig.~\ref{fig:Test_Nkol_vs_Ntr} plotting again the information backflow measured using the trace distance, its approximation using the Wigner distance and its absolute error varied over temperature $\kbT$ and coupling strength $\gamma$ at low frequency cutoff $\Omega = \omega_0$. This regime is known to show strong backflow of information \cite{Einsiedler2020,Ma-Arb_Einsiedler_2020}. Note that unlike before we now chose a linear scale instead of a logarithmic one. Again one sees that the structure of the plot using the trace distance is very well reflected in the plot using its approximation with the Wigner distance. The absolute error is now more than two decimal powers smaller than $\mathcal{N}_\mathrm{tr}$ over the whole parameter range. In general, the magnitude of information backflow in this regime is roughly two decimal power larger than in the upper subplots and accordingly, numerical fluctuations are not visible any more. Again, we see that the absolute error in the lower right plot gets significantly smaller with increasing temperature as explained above.

\section{Conclusion}\label{Conclusion}
In this paper we have extended the measure for the flow of information between an open quantum system and its environment based on the trace distance of quantum states to classical phase space models using the Kolmogorov distance between quasi-probability distributions on phase space. We explored the connection between the trace distance based measure of distinguishability of quantum states and the Kolmogorov distances for differently ordered quasi-probability distributions on phase space, and showed that for any pair of quantum states one can always find a unique optimal quasi-probability distribution for which the Kolmogorov distance coincides with the trace distance of the quantum states. We further investigated how within a limit of classical uncertainty \--- i.e. the variances in any direction of phase space become much larger than $\hbar$ \--- the Kolmogorov distances for any ordering converge to the trace distance which leads to an insightful quantum-to-classical transition for this measures of information flow and, hence, for the quantum-to-classical transition of the non-Markovianity of open systems.

We have illustrated these general results with the help of a prototypical open system model, namely the Caldeira-Leggett model of quantum Brownian motion. As expected, in this model one observes that quantum states quickly reach the limit of classical uncertainty in case of high temperatures. Quite remarkably, we have also seen that the Kolmogorov distance of the Wigner functions approximates the trace distance very well even in the range of intermediate temperatures. Our numerical results indicate that within the limit of classical uncertainty the Wigner function actually represents the distribution on phase space which yields an optimal approximation of the trace distance of quantum states. This fact is in accordance to applications of the Wigner function as semi-classical representation of the quantum density operator. Thus, at least in this limit the Wigner function Kolmogorov distance can be used to measure the information flow and the degree of non-Markovianity of the open system dynamics.

Future research might focus on the exact determination of the optimal ordering and its dependence on the system states and on the various parameters of the environment. Moreover, extensive numerical studies for different models or even a rigorous mathematical proof for the optimality of the Wigner function in the semi-classical limit would be of great interest. Finally, it would be interesting to develop possible experimental realization of the scenarios developed here to measure the information flow and its behavior in the semi-classical limit.
\begin{acknowledgments}
This work has been supported by the German Research Foundation (DFG) through FOR 5099.
\end{acknowledgments}
\appendix
\section*{Appendix}
\setcounter{section}{0}
\renewcommand{\thesection}{\Alph{section}}
\section{Representing phase space in complex plane and position-momentum coordinates}
\label{Representing phase space in complex plane and position-momentum coordinates}
Since we are aware that most people working in quantum optical phase space are familiar with a representation as complex plane where the phase space vector $\x$ is defined by a complex number $\alpha(\x)$ we will here give a little compendium of how several commonly used expressions read in the usual complex plane representation and the real phase space representation where both position and momentum are expressed in units of $\sqrt{\hbar}$ to ease comparison of our formulas with those in other publications.

For the translation from complex number to phase space vector we have
	\begin{align}
	\x(\alpha) = \sqrt{2\hbar} \begin{pmatrix} \Re(\alpha) \\ \Im(\alpha) \end{pmatrix} \quad \Leftrightarrow \quad \alpha(\x) = \frac{1}{\sqrt{2\hbar}}(q + ip).
	\end{align}
For absolute values and integration increments this implies
	\begin{align}
	|\alpha| = \frac{1}{\sqrt{2\hbar}} \|\x\|^2 \quad \und \quad d^2\alpha = \frac{1}{2\hbar} d\x
	\end{align}
With the relation between quadrature operators $\q$, $\p$ and ladder operators $\A$, $\A^\dagger$
	\begin{align}
	\X &= \begin{pmatrix} \q \\ \p \end{pmatrix} = \sqrt{\frac{\hbar}{2}} \begin{pmatrix} \A^\dagger + \A \\ i (\A^\dagger - \A) \end{pmatrix}
	\\
	\A &= \frac{1}{\sqrt{2 \hbar}} (\q + i\p)
	\end{align}
one finds for the exponents in the displacement operators we find
	\begin{align}
	-\frac{i}{\hbar} \br{\Omegabm\x}^T \X = \alpha(\x) \A^\dagger - \alpha^*(\x)\A
	\end{align}
and similarly for the kernels of the Fourier transformation
	\begin{align}
	-\frac{i}{\hbar} \br{\Omegabm\y}^T \x = \alpha(\x) \alpha^*(\y) - \alpha^*(\x)\alpha(\x).
	\end{align}
The Fourier transformation from characteristic function to quasi-probability distribution now reads
	\begin{align}
	W_\mathbb{C}(\alpha) &= \frac{1}{\pi^2} \underset{\mathbb{C}}{\int}d^2 \beta \ \chi_\mathbb{C}(\beta) \exp\Br{\alpha\beta^* - \alpha^*\beta}
	\\
	W_\Gamma(\alpha) &= \frac{1}{(2\pi\hbar)^2} \underset{\Gamma}{\int}d\y \ \chi_\Gamma(\y) \exp\Br{-\frac{i}{\hbar} \br{\Omegabm\y}^T \x},
	\end{align}
where the index $\Gamma$ or $\mathbb{C}$ indicates if it is the corresponding function in phase space or complex plane. Thus, we have the familiar expressions for Glauber P and Husimi Q-function
	\begin{align}
	&\Rho = \underset{\Gamma}{\int}d\x \ P_\Gamma (\x) \ \ketbra{\x}{\x} = \underset{\mathbb{C}}{\int}d^2 \alpha \ P_\mathbb{C}(\alpha) \ \ketbra{\alpha}{\alpha}
	\\
	&Q_\Gamma (\x) = \frac{1}{2\pi\hbar} \braket{\x|\Rho|\x} \quad \und \quad Q_\mathbb{C}(\alpha) = \frac{1}{\pi} \braket{\alpha|\Rho|\alpha}
	\end{align}
with $\ket{\x} = \ket{\alpha(\x)}$ being the same coherent state just in different notation. Accordingly one has with $\alpha = \alpha(\x)$ and $\beta = \alpha(\y)$
	\begin{align}
	\begin{split}
	\braket{\x|\y} &= \exp\Br{-\frac{1}{4\hbar}\|\x-\y\|^2} \cdot \exp\Br{-\frac{i}{2\hbar} \br{\Omegabm \y}^T \x}
	\\
	&= \exp\Br{-\frac{1}{2} |\alpha-\beta|^2} \cdot \exp\Br{\frac{1}{2}(\alpha\beta^* - \alpha^*\beta)}
	\\
	&= \braket{\alpha|\beta}.
	\end{split}
	\end{align}
\section{Transformation between differently ordered quasi-probability distributions}
\label{Transformation between differently ordered quasi-probability distributions}
We want to show, how differently ordered quasi-probability distributions are connected by convolutions with Gaussian kernels. With Eq.~(\ref{eq:D(x)}) we can define the \textit{s-ordered characteristic function} $\chi^s_\rho$ of a quantum state \cite{Serafini_2017}
	\begin{align}
	\chi^s_\rho (\y) \ &= \ \Tr{\bm{\mathcal{D}}^s(\y) \Rho} \ = \ \chi^0_\rho(\y) \cdot \exp\left[\frac{s}{4\hbar}\|\y\|^2 \right]
	\end{align}
and for two different order parameters $r$, $s$ with $r < s$ we thus get
	\begin{align}
	\begin{split}
	\chi^r_\rho (\y) \ &= \ \chi^s_\rho(\y) \cdot \exp\Br{-\frac{s-r}{4\hbar}\|\y\|^2 }
	\\
	&= \ \chi^s_\rho(\y) \cdot \epsilon^{(s-r)}(\y).
	\end{split}
	\end{align}
Eq.~(\ref{eq:T(x)}) now implies that the s-ordered quasi-probability distribution is the Fourier transformation of the s-ordered characteristic function
	\begin{align}
	\begin{split}
	W^s_\rho (\x) \ &= \F\Br{\chi^s_\rho}(\x)
	\\
	&= \ \frac{1}{(2 \pi \hbar)^2} \underset{\Gamma}{\int} d\vec{y} \ \exp\Br{ -\frac{i}{\hbar} \br{\Omegabm\vec{y}}^T \x } \cdot \chi^s_\rho(\y)
	\end{split}
	\end{align}
and thus, again for two different ordering $r$, $s$ with $r < s$ we find due to the convolution theorem
	\begin{align}
	\begin{split}
	W^r_\rho (\x) \ &= \ \F\Br{\chi^s_\rho \cdot \epsilon^{(s-r)}}(\x)
	\\
	&= \ \br{\F\Br{\chi^s_\rho} \star \F\Br{\epsilon^{(s-r)}}}(\x)
	\\
	&= \ \br{W^s_\rho \star \F\Br{\epsilon^{(s-r)}}}(\x)
	\end{split}
	\end{align}
and the Fourier transform of $\epsilon^{(s-r)}$ gives the desired Gaussian kernel $\F\Br{\epsilon^{(s-r)}}(\x) = \Gauss{}{(s-r)\hbar \1}(\x)$.
\newpage
\def\newblock{\ }
\bibliography{Bibliography_Richter-Breuer_PRA_2024}

\end{document}